\documentclass[pra,reprint]{revtex4-2}
\usepackage{amsthm}
\usepackage{amsmath}
\usepackage{latexsym}
\usepackage{amsfonts}
\usepackage{amssymb}
\usepackage{color}
\usepackage{bm,dsfont}
\usepackage{graphicx}
\usepackage{hyperref}
\usepackage{subfigure}
\usepackage{multirow}
\usepackage{xr}
\usepackage{mathrsfs}
\usepackage[noend]{algpseudocode}
\usepackage{algorithmicx,algorithm}
\usepackage{graphicx}
\usepackage{subfigure}
\usepackage{[number}
\usepackage{float}
\usepackage{svg}
\usepackage{epstopdf}

\definecolor{gray}{RGB}{123,123,123}

\theoremstyle{definition}

\newcommand{\ket}[1]{|#1\rangle} 
\newcommand{\bra}[1]{\langle#1|} 
\newcommand{\tr}[1]{\textrm{tr}\left[#1\right]} 

\graphicspath{{figures/},{pics/}}
\makeatletter

\newcommand{\Rmnum}[1]{\expandafter\@slowromancap\romannumeral #1@}
\makeatother

\begin{document}
\title{Hybrid of Gradient Descent And Semidefinite Programming for Certifying Multipartite Entanglement Structure }
\author{Kai Wu}
\affiliation{School of Science, Jimei University, Xiamen 361021,China}
\author{Zhihua Chen}
\thanks{chenzhihua77@sina.com}
\affiliation{School of Science, Jimei University, Xiamen 361021,China}
\author{Zhen-Peng Xu}
\thanks{zhen-peng.xu@ahu.edu.cn}
\affiliation{School of Physics and Optoelectronics Engineering,
Anhui University, Hefei 230601, China}
\author{Zhihao Ma}
\thanks{mazhihao@sjtu.edu.cn}
\affiliation{School of Mathematical Sciences, MOE-LSC, Shanghai Jiao Tong University, Shanghai 200240,
China}
\affiliation{Shanghai Seres Information Technology Co., Ltd, Shanghai 200040, China}
\affiliation{Shenzhen Institute for Quantum Science and Engineering, Southern University of Science and Technology, Shenzhen 518055, China}
\author{Shao-Ming Fei}
\thanks{smfei@mis.mpg.de}
\affiliation{School of Mathematical Sciences, Capital Normal University, Beijing 100048, China}
\affiliation{Max Planck Institute for Mathematics in the Sciences, 04103 Leipzig, Germany}

\begin{abstract}
Multipartite entanglement is a crucial resource for a wide range of quantum information processing tasks, including quantum metrology, quantum computing, and quantum communication. The verification of multipartite entanglement, along with an understanding of its intrinsic structure, is of fundamental importance, both for the foundations of quantum mechanics and for the practical applications of quantum information technologies. Nonetheless, detecting entanglement structures remains a significant challenge, particularly for general states and large-scale quantum systems. To address this issue, we develop an efficient algorithm that combines semidefinite programming with a gradient descent method. This algorithm is designed to explore the entanglement structure by examining the inner polytope of the convex set that encompasses all states sharing the same entanglement properties. Through detailed examples, we demonstrate the superior performance of our approach compared to many of the best-known methods available today. Our method not only improves entanglement detection but also provides deeper insights into the complex structures of many-body quantum systems, which is essential for advancing quantum technologies.
\end{abstract}

\maketitle

\section{Introduction}

Multipartite quantum entanglement plays a crucial role in quantum communication and information processing tasks, including quantum computing, quantum cryptography, and quantum metrology \cite{PhysRep, Geza, Liu_2020, nature629}.
Quite different from the bipartite case, multipartite quantum entanglement exhibits a rich structure.a much richer structure. Two key concepts have been introduced to characterize this structure: entanglement producibility and entanglement partitionability \cite{Anders,Otfried}.  Intuitively, for pure states, entanglement partitionability refers to the number of subsystems that can be separated from each other in the multipartite system, while entanglement producibility denotes the size of the largest entangled subsystem. Recently, the concept of entanglement stretchability has been introduced \cite{Szalay2019}, which stands for the difference between the entanglement producibility and entanglement partitionability. {Besides, the general theory of one-parameter families of partial entanglement properties and the resulting entanglement depth-like quantities have been proposed. The physically meaningful entanglement structure such as squareability, toughness and the degree of freedom have been constructed \cite{2408}.}

Quantum entanglement structure has a wide range of applications, including in quantum networks, quantum metrology, and quantum phase transitions \cite{nature629, Geza,Liu_2020, Zhihong,Li_2024,2408}. However, as the number of subsystems grows, verifying the entanglement structure becomes a challenging problem \cite{Anders,Otfried}. While one could, in theory, detect the entanglement structure through state tomography, this approach becomes impractical due to the exponential growth of the underlying Hilbert space dimensions. It is important to note that quantum states sharing the same entanglement structure form a convex set. Significant efforts have been made to develop practical techniques that explore the outer polytopes of this convex set. These efforts include the use of more accessible inequalities derived from spin-squeezing, the entries of density matrices, local uncertainty relations, Wigner-Yanase skew information, Fisher information, and even machine learning approaches \cite{PhysRep,Taming,Huber,Bae,Graph,Jens,Hong_2015,Chen_2018,Hong_2016,Gao_2014,Hong_2021,Vitagliano_2011,Chen_2005,Hyllus_2012,Geza,Gessner,Hong_2021,Chen_2021,Zhou,ChenJiYao_2016,Bancal_2011,Collins_2002,Aloy_2019,LinPeiSheng_2019,LiangYeongCherng_2015,Tura_2019,Lewenstein_2022,Shen_2020, Lepori_2023,Guo_2022}.

Nevertheless, most techniques for verifying entanglement structures through the investigation of the outer polytope only provide sufficient conditions for identifying entanglement structures outside the convex set. For instance, all bi-separable states form a convex set, and corresponding witnesses provide a sufficient condition for detecting genuine multipartite entanglement. However, to establish a necessary condition for genuine multipartite entanglement, it is crucial to verify bi-separability from the inner polytope of the convex set. This requires providing sufficient conditions for separability.

But verifying separability is proved to be a difficult problem. In contrast to the techniques used to detect entanglement structure from the outer polytope, only a few progress has been made toward exploring the inner polytope of the convex set. Semidefinite programs has been employed to certify separability \cite{PhysRevA.70.062309,PhysRevLett.103.160404}. Gilbert's algorithm combined with gradient methods has been utilized to address the convex membership problem \cite{PhysRevLett.120.050506}. Additionally, truncated moment sequences and the inequalities based on the Bloch representations of the quantum states have been applied to tackle the problem of separability \cite{SciRep8.1442, PhysRevA.96.032312}.
An algorithm based on neural networks has been developed to approximate separable states \cite{PhysRevResearch.4.023238} and a variational quantum algorithm has been proposed for the verification of separability \cite{PhysRevA.106.062413}.
Recently, an algorithm based on the adaptive polytope approximation has been proposed to certify the quantum separability of bipartite and multipartite quantum systems~\cite{ohst2023certifying}. {Each iteration of this algorithm involves solving a semidefinite programming (SDP) problem. The algorithm iteratively refines each polytope to converge to a satisfactory outcome. Its performance relies on the well-chosen initial polytopes to ensure an adequately accurate result.}

The aforementioned methods are primarily focused on certifying the separability of quantum states, also known as quantum partitionability. However, there are few approaches dedicated to addressing the issue of quantum producibility, { strethability, toughness and squareability, et al}. Drawing inspiration from the algorithm described in \cite{ohst2023certifying}, we develop a novel algorithm designed to explore the entanglement structure of multipartite states, encompassing quantum partitionability, producibility, {toughness and squareability, et al. Our proposed algorithm commences with random initial polytopes and refines them by using the gradient descent method to yield certain favorable polytopes. These optimized polytopes are subsequently employed as the starting point for a semidefinite program that aims to detect the entanglement structure. A key advantage of our new algorithm is its reduced sensitivity to the initial choice of the random polytopes, ensuring that it can converge to a satisfactory result with greater reliability and less dependence on the initial conditions.}

\section{PRELIMINARY}

A $n$-partite pure state $\ket{\psi}$ is $h$-producible if $\ket{\psi}=\otimes_{i}\ket{\psi_{X_i}}$ with $\ket{\psi_{X_i}}$ being at most a $h$-partite entangled state. The state is $k$-partitionable if $\ket{\psi}=\otimes_{i=1}^k\ket{\psi_{X_i}}.$ Here we call $\gamma=\{X_i\}$ a partition of $\mathcal{N}=\{12\cdots n\}$ which satisfies $\mathcal{N}=\cup_i X_i$ and $X_i\cap X_j=\emptyset$. We denote $|S|$ the size of $S$, then $\ket{\psi}$ is said to be
1) $h$-producible if $\max |X_i|\le h$ for all $i$; 2) $k$-partitionable if the number of the parts in the partition, $|\gamma|=|\{X_i\}| \le k$.

\begin{figure}
\includegraphics[width=1.0\columnwidth]{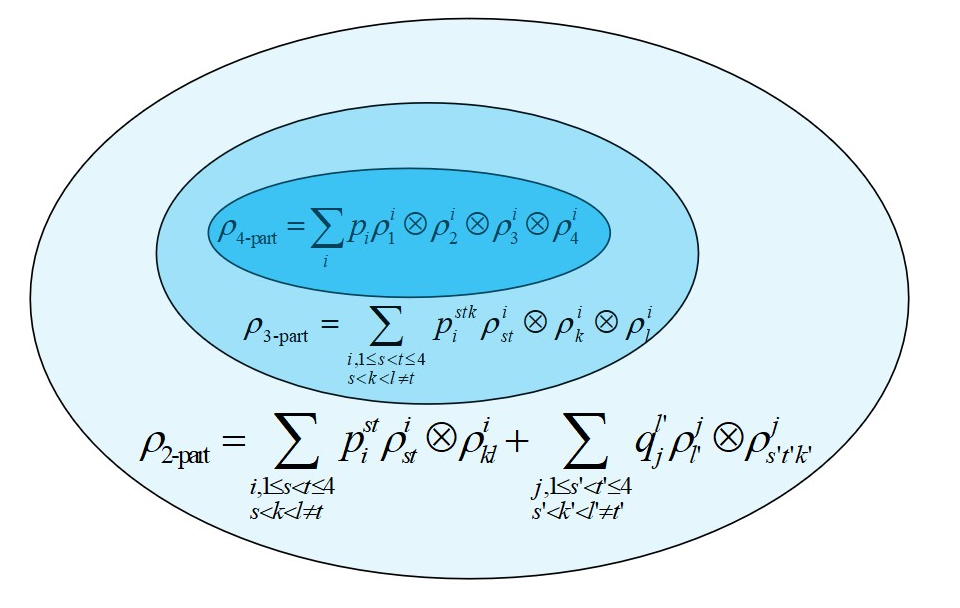}
\caption{\label{4-part} The partitionability of the $4$-partite state.}
\end{figure}

\begin{figure}
\includegraphics[width=1.0\columnwidth]{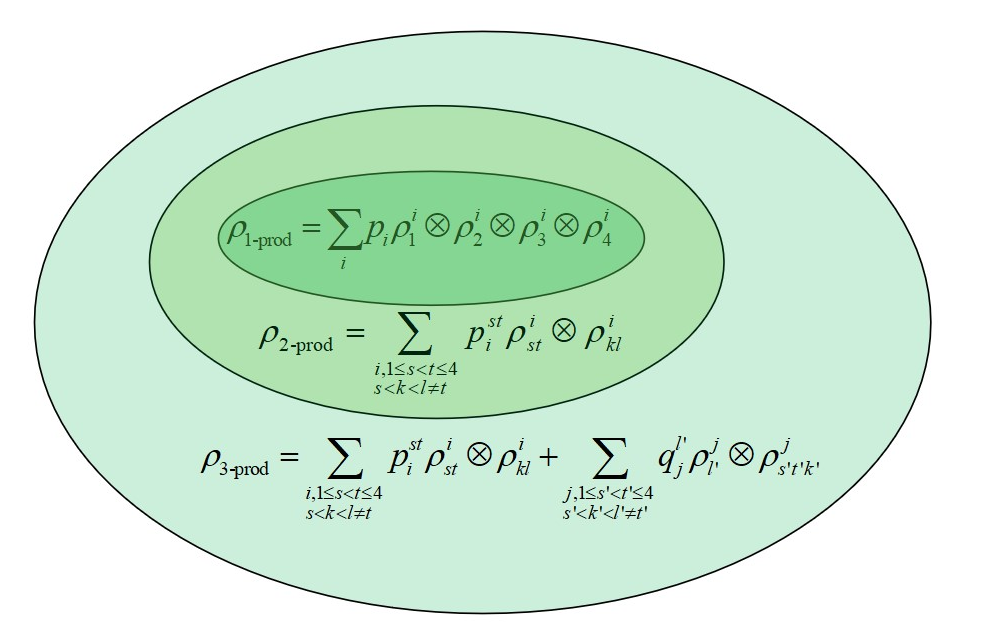}
\caption{\label{4-prod} The producibility of the $4$-partite state.}
\end{figure}

A mixed state $\rho$ is called $h$-producible ($k$-partitionable) if it can be decomposed as a convex combination of $h$-producible ($k$-partitionable) pure states. As an example, the partitionability and producibility of the $4$-partite states are illustrated in Fig.\ref{4-part} and \ref{4-prod}, respectively. The $h$-producible or $k$-partitionable pure states have different types. For the $4$-partite quantum states, $\{X_i\}$ has seven different types of partitions, ($\{123,4\}$, $\{124,3\}$, $\{134,2\}$, $\{234,1\}$,
$\{12,34\}$, $\{13,24\}$ or $\{14,23\}$) for 3-producibility.
{We denote these partitions as 3|1 or 2|2 for convenience, where the numbers 3,1 and 2 represent the number of the parties in the subsystems.}. To verify whether a mixed state is $k$-partitionable or $h$-producible, one needs to find the decompositions which contain different types of $k$-partitionable or $h$-producible pure states. An semidefinite program (SDP) based method was developed to create evolving polytopes, whose nodes are different kinds of separable states, for confirming the quantum partitionability~\cite{ohst2023certifying}. For example, consider the state $\rho(t)=t \rho+\frac{1-t}{d}\mathbb{I}_d$, where $d$ is the dimension of $\rho$ and $\mathbb{I}_d$ is the identity matrix. To check whether $\rho$ is bi-separable, SDP can be utilized to maximize $t$ such that $\rho(t)$ is bi-separable by generating a set of initial states $\{ \varrho_i \}$ randomly and optimizing $\{\tau_i\}$,
\begin{align} \label{SDP}
\begin{aligned}
\max &\ \ t \\
\mathrm{w.r.t.} &\ \ t, \tau_i \succcurlyeq 0 \\
\mathrm{s.t.} &\ \ \rho(t) = \sum_i \varrho_i \otimes \tau_i,
\end{aligned}
\end{align}
where the trace of $\tau_i$ is not necessarily one. Then maximizing $t$ such that $\rho(t)$ is still a bi-separable state
by using SDP with $\{\tau_i\}$ (obtained from the last step) being the initial states and $\varrho_i$ being optimized.
$\rho$ is a bi-separable state if $t=1$ and we can obtain the maximum of $t$ such that $\rho(t)$ is bi-separable.

\section{Detection of entanglement structure}

The performance of the SDP-based algorithm in \cite{ohst2023certifying} depends on the choice of the initial polytope. As the complexity increases with the increasing system, inspired by the previous works \cite{ohst2023certifying,PhysRevLett.120.050506,PhysRevA.96.032312}, we propose a systematic approach in the following to select the initial polytope so as to guarantee the efficiency and performance.

\begin{figure}
\includegraphics[width=1.0\columnwidth]{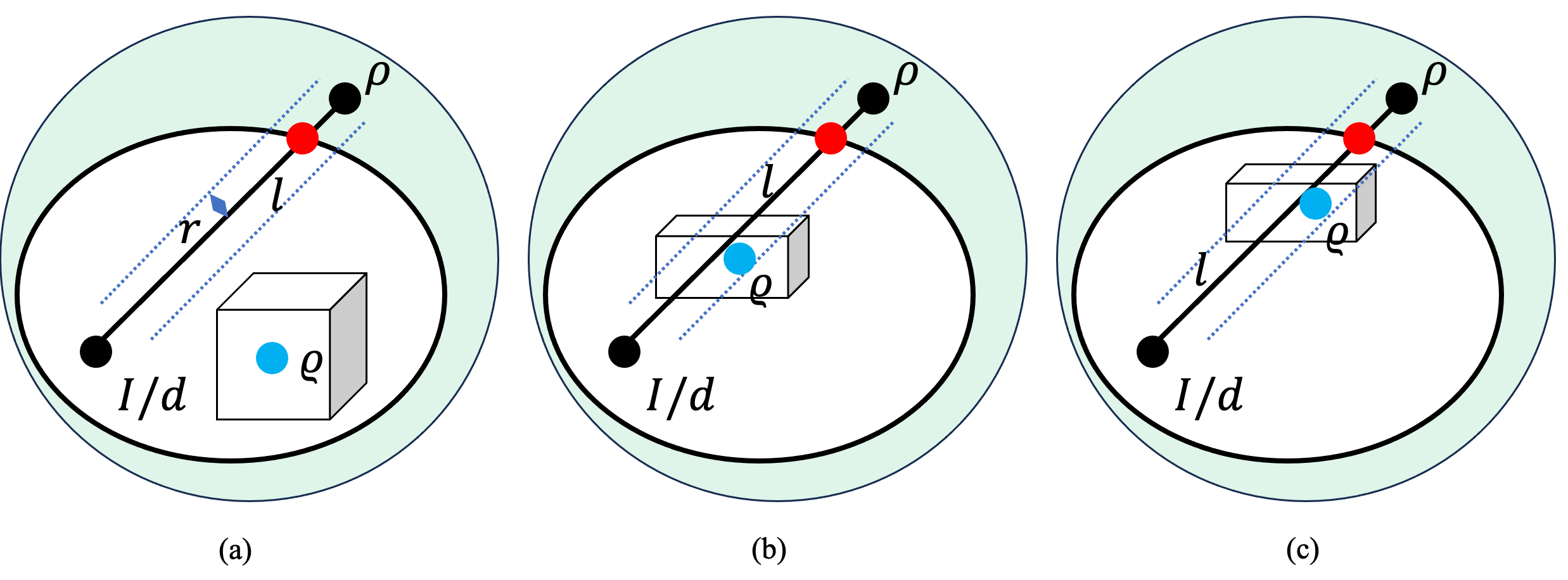}
\caption{\label{fig:next-algo} (a) Initialise an arbitrary polytope $\mathcal{P} = \{ \varrho_i \}$ and $\varrho = \sum_i P_i \varrho_i$. (b) Minimize the distance between the quantum state $\varrho$ and the segment $l$. (c) Minimize the distance between the quantum state $\varrho$ and $\rho$.}
\end{figure}

{We introduce a new algorithm to certify the entanglement structure of the mixed state {$\rho(t) = t \rho + (1 - t) \mathbb{I}_d / d$} for given $\rho$. Especially, we are interested in the maximum of $t$ such that $\rho(t)$ still has the targeted entanglement structure. Our algorithm involves two stages as illustrated in Fig.~\ref{fig:next-algo}.} { Different from \cite{ohst2023certifying}, initially, we employ gradient descent at the first stage},
which is followed by SDP at the second stage. The gradient descent is conducted within the Pytorch framework, while the SDP is facilitated through the PICOS \cite{PICOS} library. The whole procedure can be summarized as follows.

\begin{enumerate}
{\item Initialize an arbitrary polytope $\mathcal{P}$ determined by a set of states $\{ \varrho_i \}$ with a certain entanglement structure, namely, either $h$-producibility or $k$-partitionability, and a set of probabilities $\{ p_i \}$ with $\sum_i p_i = 1$.
An example is illustrated in Fig. \ref{fig:next-algo}(a).}

{\item Employing the gradient descent algorithm, we minimize the geometric distance from the state $\varrho = \sum_i p_i \varrho_i$ to the line segment $l$ between $\rho$ and the maximally mixed state, as shown in Fig. \ref{fig:next-algo}(b). A predefined threshold $r$ serves as the criterion for progression to the next step, ensuring a systematic approach to state optimization.}

{\item Using gradient descent to minimize the geometric distance between the states $\varrho$ and $\rho$, ensuring that the polytope's boundaries progressively converge to $\rho$, as illustrated in Fig. \ref{fig:next-algo}(c). The optimization process is directed by a loss function $\lVert \rho - \varrho \rVert = \sqrt{\tr{(\rho - \varrho )^{\dag} (\rho - \varrho)}}$, with the initial polytope configuration determined by the results of step 2. If the distance, as defined in step 2, exceeds the predetermined threshold $r$, the algorithm reverts to step 2 to ensure that the distance between $\varrho$ and the line segment $l$ remains within the specified limit.}

\item Finally, we use SDP to approximate the optimal $t_0$. The initial points are chosen as the results in step 3. We choose one subsystem from each vertex of $\mathcal{P}$ as the unknown one, which will be optimized using SDP. The rest subsystems from each vertex are denoted as a new set $\mathcal{P}^{\prime}$.
For example, to investigate the $3$-producibility of a $4$-partite state, we have $\mathcal{P}=\{\varrho_1\otimes \tau_{234},\varrho_2\otimes \tau_{134},\varrho_3\otimes \tau_{124},\varrho_4\otimes \tau_{123},\varrho_{12}\otimes \tau_{34},\varrho_{13}\otimes \tau_{24},\varrho_{14}\otimes \tau_{23}\}$ and
$\mathcal{P}^{\prime}=\{\varrho_1,\varrho_2,\varrho_3,\varrho_4,\varrho_{12},
\varrho_{13},\varrho_{14}\}$.
We do this iteratively for each subsystem. The details of the SDP are as follows:
\begin{align} \label{SDP}
\begin{aligned}
\max &\ \ t \\
\mathrm{w.r.t.} &\ \ t, \tau_i \succcurlyeq 0 \\
\mathrm{s.t.} &\ \ \rho(t) = \sum_{\varrho_i \in \mathcal{P}^{\prime}} \varrho_i \otimes \tau_i,
\end{aligned}
\end{align}
where the trace of $\tau_i$ is not necessarily one.
\end{enumerate}

{While step 2 permits the convergence of $\varrho$ towards the line segment $l$, the intersection of the polytope $\mathcal{P}$ with $l$, and consequently the feasibility of SDP, cannot be assured. To improve this, we optimize the quantum subsystems corresponding to the largest part of each node via SDP. For fully separable states and certain entanglement structures for which the SDP is infeasible, we can also construct an initial polytope by using only states derived from mutually-unbiased bases (MUBs), which not only covers a substantial volume but also encapsulate the maximally mixed states. In particular, for the case that the local dimension is $2$, the constraint condition in Eq.~\eqref{SDP} is transformed into the following one:}
\begin{equation*}
    \rho(t) = \sum_{\varrho_i \in (\mathcal{P}^{\prime} \bigcup \mathcal{P}_{N-1})} \varrho_i \otimes \tau_i,
\end{equation*}
where ${\cal P}_{N-1}$ is given by
\begin{equation}
  \big\{ \bigotimes_{i=1}^{N-1} \rho_i, \rho_i \in \big\{ \frac{(\mathbb{I}_2 \pm \sigma_x)}{2}, \frac{(\mathbb{I}_2 \pm \sigma_y)}{2}, \frac{(\mathbb{I}_2 \pm \sigma_z)}{2} \big\} \big\}.
\end{equation}
Since $\mathcal{P}_{N-1}$ contains the maximally mixed state, the feasibility of SDP can be guaranteed.

\section{{Applications of the algorithm to detect entanglement structure}}
Firstly, we study the full separability of the states $\rho(t)=t\rho+(1-t)\mathbb{I}_d / d$, where $\rho$ is any state up to six qubits. The results are computed by the algorithm based on the inner polytope. We compare the results with the known ones in Table \ref{tab:full}.
\begin{table}[h]
\caption{\label{tab:full} Comparison of the { maximum of $t$ such that the state $\rho(t)$ remains} fully separable. The maximum of $t$ is computed using our method and the ones by adaptive polytopes \cite{ohst2023certifying}. The ``-'' in the table signifies that \cite{ohst2023certifying} either did not provide the results or the code provided by \cite{ohst2023certifying} was unable to perform the calculations.
}
\begin{ruledtabular}
\begin{tabular}{lcc}
\textrm{Quantum States}&
\textrm{Our result}&
\textrm{Lower bound of \cite{ohst2023certifying}}\\
\colrule
GHZ(6 Qubits) & 0.0303 & -\\
W(6 Qubits) & 0.0235 & -\\
Cluster state (6 Qubits) & 0.0303 & -\\
GHZ(5 Qubits) & 0.05882 & 0.05878 \\
W(5 Qubits) & 0.0471 & 0.0470 \\
Cluster state (5 Qubits) & 0.0588 & - \\
GHZ(4 Qubits) & 0.1111 & 0.1111\\
W(4 Qubits) & 0.0926 & 0.0926\\
Cluster state (4 Qubits) & 0.1111 & 0.1111\\
Dicke (4 Qubits, 2 ex.) & 0.0857 & 0.0857\\
\end{tabular}
\end{ruledtabular}
\end{table}

For the $4$ qubit, $5$-qubit and $6$-qubit noisy GHZ states, the results computed by our algorithm based on the inner polytope give the necessary and sufficient values $1/9$, $1/17$ and $1/33$ of $t$ for the full separability, respectively \cite{Chen_2018}. {Since we optimize the initial polytopes for SDP by using the gradient descent algorithm and the polytopes containing maximally mixed state, our algorithm performs better for some cases. For example,  our method gives rise to the maximum of noise for the full-separability of 6-qubit GHZ state, 6-qubit W state, 5-qubit and 6-qubit cluster state mixed with white noise, while the algorithm in \cite{ohst2023certifying} fails.}

Then, we utilize our algorithm based on the inner polytope to compute the entanglement parititionability and the entanglement producibility of $\rho(t)$ when $\rho$ is 4-qubit W state, 4-qubit and 5-qubit GHZ states, see Tables \ref{tab:w-4-inside}, \ref{tab:ghz-4-inside} and \ref{tab:ghz-5-inside}, respectively. We compare our results with the best ones known to date as follows.
\begin{table}[h]
\caption{\label{tab:w-4-inside} Comparison of {the maximum of $t$ such that $\rho(t)$ remains $k$-partitionable  or $k$-producible. The maximum of $t$ is} computed by our algorithm and the best known ones to date in the literature for the mixture of 4-qubit W state and white noise.
}
\begin{ruledtabular}
\begin{tabular}{ccc}
\textrm{Entanglement Structure}&
\textrm{Our result}&
\textrm{Other known results}\\
\colrule
3-part & 0.247 & 0.247\cite{PhysRevResearch.4.023238} \\
2-prod & 0.247 & 0.245\cite{PhysRevLett.120.050506} \\
2-part & 0.471 & 0.474\cite{Taming} \\
\end{tabular}
\end{ruledtabular}
\end{table}
\begin{table}[h]
\caption{\label{tab:ghz-4-inside} Comparison of {the maximum of $t$ such that $\rho(t)$ remains $k$-partitionable  or $k$-producible. The maximum of $t$ is} computed by our algorithm and the known ones in the literature for the mixture of 4-qubit GHZ state and white noise.
}
\begin{ruledtabular}
\begin{tabular}{ccc}
\textrm{Entanglement Structure}&
\textrm{Our result}&
\textrm{Other known results}\\
\colrule
3-part & 0.200 & 0.198\cite{PhysRevLett.120.050506,PhysRevResearch.4.023238} \\
2-prod & 0.273 & 0.271\cite{PhysRevResearch.4.023238} \\
2-part & 0.465 & 0.467\cite{Taming}(exact) \\
\end{tabular}
\end{ruledtabular}
\end{table}
\begin{table}[h]
\caption{\label{tab:ghz-5-inside} Comparison of {the maximum of $t$ such that $\rho(t)$ remains $k$-partitionable  or $k$-producible. The maximum of $t$ is} computed by our algorithm and the known ones in the literature for the mixture of 5-qubit GHZ state and white noise.
}
\begin{ruledtabular}
\begin{tabular}{ccc}
\textrm{Entanglement Structure}&
\textrm{Our result}&
\textrm{Other known Results}\\
\colrule
4-part & 0.094 & 0.094 \cite{Chen_2018}\\
2-prod & 0.238  & 0.19\cite{PhysRevLett.120.050506} \\
3-part & 0.238  & 0.238\cite{Chen_2018}\\
3-prod & 0.385 & 0.278\cite{PhysRevLett.120.050506} \\
2-part & 0.484 & 0.484\cite{Chen_2018}\\
\end{tabular}
\end{ruledtabular}
\end{table}

{ Table \ref{tab:w-4-inside} shows that our results based on inner polytope perform better than the best one so far in \cite{PhysRevLett.120.050506} in the case of $2$-producibility for $4$-qubit noisy W state. Table \ref{tab:ghz-4-inside} shows that the bound corresponding to $3$-partitionablility is $0.200$ for $4$-qubit noisy GHZ state, which coincides with the necessary and sufficient value $p=0.2$ \cite{Chen_2018}. Besides, the bound $0.273$ for $2$-produciblity is better than the best one so far in \cite{PhysRevLett.120.050506}}.

{Table \ref{tab:ghz-5-inside} is for $5$-qubit noisy GHZ state. Our results of $2/3/4$-partitionablity coincide with the optimal known values given in \cite{Chen_2018}. For the producibility, our results are better compared with the ones in \cite{PhysRevLett.120.050506}. From \cite{Hong_2021} the exact bound is no more than $0.238$ for $2$-producibility of $\rho(t)$, which is identical to our bound $0.238$.}

{For the entanglment structure proposed in \cite{2408}, we compute the maximum of $t$ such that $\rho(t)$ maintains the toughness of $k$ and the squareability of $k$ in Table \ref{tab:ghz-cluster-tou-squ}. Here, toughness-1 (2) corresponds to an entanglement structure characterized by partitions 4|1 (3|2). Squareability values of 7, 9, 11, 13, and 17 correspond to entanglement structures with partitions 2|1|1|1, 2|2|1, 3|1|1, 3|2, and 4|1, respectively. Notably, for the noisy GHZ state, the results for toughness-2 align with those for partitionability 2. Similarly, the results for squareability-9 and 13 align with those for producibility 2 and 3, respectively.}

{
\begin{table}[h]
\caption{\label{tab:ghz-cluster-tou-squ} The maximum of $t$ such that $\rho(t)$ remains the toughness of $k$ or Squareability of $k$ for 5-qubit GHZ states or 5-qubit cluster state mixed with white noise.
}
\begin{ruledtabular}
\begin{tabular}{ccc}
\textrm{Ent. Str.}&
\textrm{noisy cluster state}& \textrm{noisy GHZ state}\\
\colrule
toughness-1 & 0.273 & 0.238\\
toughness-2 & 0.360 & 0.484\\
squareablity-7 & 0.111 & 0.094\\
squareablity-9 & 0.158 & 0.238\\
squareablity-11 & 0.210 & 0.238\\
squareablity-13 & 0.272 & 0.385\\
squareablity-17 & 0.360 & 0.484\\
\end{tabular}
\end{ruledtabular}
\end{table}
}

{It is observed in the above computation that the increase of the number of iterations in gradient descent method corresponds to a greater value of $t$.  Consequently, a sufficient number of iterations in the gradient descent method is needed to ensure the optimal results during the optimization process. For example, detecting the entanglement structure of a 5-qubit GHZ state mixed with white noise requires approximately 5000 iterations in the gradient descent method. We have thus set the maximum number of iterations to 5000. Typically, within these 5000 iterations, the difference between consecutive results becomes sufficiently small, allowing us to consider the outcome as a reasonably accurate result. In contrast, for the detection of entanglement structures presented in Tables \ref{tab:full}, \ref{tab:w-4-inside} and \ref{tab:ghz-4-inside}, a maximum of only 1000 iterations is required. This difference in the number of iterations underscores the complexity and varying demands of entanglement detection across different quantum states. However, it should be noted that this approach does not guarantee obtaining the global minimum for geometric distances in steps 2 and 3. Uncertainties and limitations remain, and further research is needed to enhance the accuracy and reliability of finding the optimal solution.}

Additionally, our algorithm can provide the exact decomposition of $k$-partitionable ($h$-producible) states. {However, despite potentially not capturing the global optimum for the geometric distances in steps 2 and 3, our algorithm can still yield a promising value of $t$ that guarantees a high-fidelity between the decomposition of $\rho(t)$ obtained by our algorithm  and the original $\rho(t)$.
Calculations show that the fidelity between the state $\rho(t)$ and the decomposition produced by our algorithm reaches an impressive $99.999\%$. This high fidelity confirms the accuracy and reliability of our algorithm, making it a robust tool for detailed analysis of quantum entanglement structures. The near-perfect fidelity indicates that the algorithm's output closely matches the actual state’s decomposition, thereby confirming the high precision of our method in practical applications.}

Furthermore, we can detect the entanglement structure of any state $\rho$ mixed with any fully separable states. For example, let us consider $\rho(t)=t \ket{W}\bra{W}+(1-t)\rho_{f}$, where $\ket{W}=(\ket{0001}+\ket{0010}+\ket{0100}+\ket{1000})/2$ and $\rho_{f}=\bigotimes\limits_{i=1}^4\tau_i$ with $\tau_i=(3|0\rangle\langle 0|+|1\rangle\langle 1|)/4$. Our algorithm shows that $\rho(t)$ is fully separable when $t<0.1468$ and 2-partitionable if $t<0.58$, while it was shown in \cite{PhysRevLett.120.050506} that $\rho(t)$ is fully separable when $t<0.1464$ and 2-partitionable for $t<0.5623$.

To guarantee the universality of our algorithm, the initial states in our algorithm are optimized by the gradient conjugate method. In our approach, we focus on optimizing the states of one subsystem, with the rest subsystems' states being initialized accordingly. In this way, the runtime of our algorithm could be slightly longer than that given in \cite{ohst2023certifying}. By optimizing the states of a single subsystem, we ensure that the entanglement properties are accurately captured, while the computational feasibility of the algorithm is maintained. Although this approach may increase the computational time, it is a necessary trade-off to achieve the high level of precision and universality offered by our algorithm. The additional runtime is a small price to pay for the enhanced reliability and broader applicability of our method in certifying multipartite entanglement structures. {Take 5-qubit GHZ state mixed with white noise as the example, we list the time consumption for computing the maximum of $t$ such that $\rho(t)$ remains certain entanglement structure in Table\ref{tab:time-consump}. We use 100 quantum states for each partition in the decompositions  to construct the polytope $\mathcal{P}$. The epoch of the gradient descent is 1000. The memory and the CPU of the computer is 16 GB and 6-core, 12-thread processor respectively.}

{
\begin{table}[h]
\caption{\label{tab:time-consump} Time consumption for computing the maximum of $t$ such that $\rho(t)$ remains certain entanglement structure for 5-qubit GHZ states mixed with white noise. The unit is seconds.
}
\begin{ruledtabular}
\begin{tabular}{ccc}
\textrm{Ent. Str.}&
\textrm{GD}& \textrm{SDP}\\
\colrule
toughness-1 & 177  & 68 \\
toughness-2 & 504 & 231\\
producibility-2 & 750 & 133\\
producibility-3 & 354 & 111\\
partitionability-2 & 514 & 240\\
partitionability-3 & 1263 & 422\\
\end{tabular}
\end{ruledtabular}
\end{table}
}
{For the entanglement structure detection of 5 or fewer qubit quantum states, the SDP algorithm takes less time than the gradient descent method. However, for the 6-qubit noisy states with toughness 1, the SDP algorithm takes more time than the gradient descent method, with time of 150 seconds and 44 seconds, respectively. When SDP algorithm is employed to detect other entanglement structures for 6-qubit quantum states, the program may encounter a crash due to insufficient memory allocation because of the excessive number of partitions.}

{In quantum metrology, the entangled states are used to enhance measurement precision, as illustrated by the inequality $\frac{1}{n}F_Q(\rho,J_z)\leq D(\rho)$. Here $D(\rho)$ represents the entanglement depth of the state $\rho$, while $F_Q(\rho,J_z)$ denotes the quantum Fisher information, which serves as a key figure of merit for characterizing the precision of metrology. Recently, a more stringent bound has been introduced for $F_Q(\rho,J_z)$, $F_Q(\rho,J_z)\leq D_{sq}(\rho)\leq D_{sq}^{oF}(\rho)$ \cite{2408}. The quantity $D_{sq}(\rho)$  refers to the squareability of entanglement. Additionally, $D_{sq}^{oF}(\rho)$ corresponds to $n$ times the average size of the entangled subsystems. Our algorithm is able to approximate these entanglement based quantities for arbitrary quantum states, thereby aiding in the estimation of metrological precision.
The structure of multipartite entanglement directly influences the functionality and efficiency of quantum networks. For instance, the type and degree of entanglement determine the capacity of a quantum network to transmit information, the robustness of the network against noise and errors, and the complexity of the quantum gates that can be implemented. In particular, certain multipartite entanglement states, such as GHZ states or cluster states, are highly sought after for their potential in realizing quantum error correction and fault-tolerant quantum computation. Our algorithm has the potential of detecting the entanglement structure within quantum networks and quantum error correction, providing valuable insights into their performance in quantum information processing.}

{Our algorithm is capable of calculating the entanglement structure of quantum states that result from the convex combination of any quantum state and a separable state, including structures such as entanglement partitionability and producibility. Due to the inherent complexity of  entanglement structures, directly applying the SDP algorithm to detect the entanglement structure may not always yield a solution. To address this issue, we use a gradient descent algorithm to search for quantum states with the desired entanglement structure, ensuring that they are close to the target state $\rho(t)$. These states are then used as the initial points for the SDP. This approach allows us to obtain accurate results even for entanglement structures that were previously difficult to detect.}

{However, with increase of the system's dimensionality, the computational resources required also grow. For example, the SDP algorithm may demand substantial memory, leading to longer computation times. To alleviate this complexity, more efficient SDP solvers, such as quantum-inspired algorithms, exploitation of problem-specific structures and techniques like parallelization and distributed computing, could be employed to reduce the computational cost of detecting entanglement structures.}

\section{Conclusion}
By leveraging the property that all quantum states with the same entanglement structure form a convex set, we have developed an algorithm that combines SDP with the gradient descent method to approximate this convex set from within. The gradient descent method is used to optimize the initial points for the SDP, increasing the likelihood of reaching the global optimum. Our algorithm surpasses some of the best-known methods for characterizing multipartite entanglement structures and is more versatile in exploring various types of entanglement.  This approach has potential applications in a variety of   quantum information tasks such as  quantum metrology and quantum networks.

\medskip
\noindent{\bf ACKNOWLEDGEMENTS}\, \, This work is supported by the Fundamental Research Funds for the Central Universities; the National Natural Science Foundation of China (NSFC) under Grants  12071179, 12305007, 12371132, 12075159 and 12171044; Anhui Provincial Natural Science Foundation (Grant No.
2308085QA29); the Alexander von Humboldt Foundation;  Natural Science Foundation of Shanghai (Grant No. 20ZR1426400); Shenzhen Institute for Quantum Science and Engineering, Southern University of Science and Technology (Grant Nos. SIQSE202005); the specific research fund of the Innovation Platform for Academicians of Hainan Province.

\bibliography{biblio}
\end{document}